\documentclass[prd%
               ,showpacs%
               ,floatfix%
               ,preprintnumbers%
               ,superscriptaddress%
           ,aps
               ,amsmath%
               ,amssymb%
           ,nofootinbib
               ]{revtex4}
\usepackage[dvips]{color,graphicx}% Include figure files
\usepackage{dcolumn}% Align table columns on decimal point
\usepackage{bm}% bold math
\usepackage{amsmath}
\usepackage{ulem}
\definecolor{ref}{rgb}{1.0,0.0,0.0}
  \makeatletter
  \def\mathcomposite{%
     \@ifstar
        {\def\@mathcomposite@option{%
            \baselineskip\z@skip\lineskiplimit-\maxdimen}%
         \@mathcomposite}%
        {\let\@mathcomposite@option\offinterlineskip
         \@mathcomposite}}
  \def\@mathcomposite{%
     \@ifnextchar[\@@mathcomposite{\@@mathcomposite[0]}}
  \def\@@mathcomposite[#1]#2#3#4{%
     #2{\mathchoice
        {\@mathcomposite@{#1}{#3}{#4}\displaystyle{1}}%
        {\@mathcomposite@{#1}{#3}{#4}\textstyle{1}}%
        {\@mathcomposite@{#1}{#3}{#4}%
         \scriptstyle\defaultscriptratio}%
        {\@mathcomposite@{#1}{#3}{#4}%
         \scriptscriptstyle\defaultscriptscriptratio}}}
  \def\@mathcomposite@#1#2#3#4#5{%
     \vcenter{\m@th\@mathcomposite@option
        \dimen@\f@size\p@\dimen@#1\dimen@\dimen@#5\dimen@
        \divide\dimen@ 18
        \edef\@mathcomposite@skipamount{\the\dimen@}%
        \ialign{\hfil$#4##$\hfil\cr
           #2\crcr
           \noalign{\vskip\@mathcomposite@skipamount}%
           #3\crcr}}}
  \makeatother

\newcommand{\Sqrt}[1]{\sqrt{\mathstrut #1}}
\def\bfm#1{\mbox{\boldmath $#1$}}
\def\bfsm#1{\mathstrut\mbox{\scriptsize{\boldmath $#1$}}\mathstrut}

\newcommand{\Eqn}[1]{Eq.~(\ref{#1})}  % includes ``Eq.'' in front

\newcommand{\beq}{\begin{equation}}
\newcommand{\eeq}{\end{equation}}
\newcommand{\ba}{\begin{array}}
\newcommand{\ea}{\end{array}}
\newcommand{\bea}{\begin{eqnarray}}
\newcommand{\eea}{\end{eqnarray}}
\newcommand{\bal}{\begin{align}}  % align in amsmath is better than eqnarray
\newcommand{\eal}{\end{align}}
\newcommand{\bi}{\begin{itemize}}  %\setlength{\itemsep}{0\parsep}}
\newcommand{\ei}{\end{itemize}}
\newcommand{\ben}{\begin{enumerate}}  %\setlength{\itemsep}{0\parsep}}
\newcommand{\een}{\end{enumerate}}

\newcommand\hide[1]{}

\newcommand{\tr}{\mbox{tr}}

\newcommand{\Det}{\mbox{Det}}

\newcommand{\ie}{{i.e.}}

\newcommand{\ds}[1]{
  \setbox0=\hbox{\ensuremath{#1}}
  \hbox to\wd0{\hbox to0pt{\hbox to\wd0{\hss/\hss}\hss}\box0}}

\newcommand{\MeV}{\,{\rm MeV}}

\newcommand{\diag}{{\rm diag}}

\begin{document}
\preprint{BARI-TH/08-588}
\title{Enforced neutrality and color-flavor unlocking\\
in the three-flavor Polyakov-loop NJL model}

\author{H. Abuki}\email{hiroaki.abuki@ba.infn.it}
\affiliation{I.N.F.N., Sezione di Bari, I-70126 Bari, Italy}
\author{M. Ciminale}\email{marco.ciminale@ba.infn.it}
\affiliation{I.N.F.N., Sezione di Bari, I-70126 Bari, Italy}
\affiliation{Universit\`a di Bari, I-70126 Bari, Italy}
\author{R. Gatto}\email{raoul.gatto@physics.unige.ch}
\affiliation{D\'epartement de Physique Th\'eorique, Universit\'e
de Gen\`eve, CH-1211 Gen\`eve 4, Switzerland}
\author{G. Nardulli}\email{giuseppe.nardulli@ba.infn.it}
\affiliation{I.N.F.N., Sezione di Bari, I-70126 Bari, Italy}
\affiliation{Universit\`a di Bari, I-70126 Bari, Italy}
\author{M. Ruggieri}\email{marco.ruggieri@ba.infn.it}
\affiliation{I.N.F.N., Sezione di Bari, I-70126 Bari, Italy}
\affiliation{Universit\`a di Bari, I-70126 Bari, Italy}

\date{\today}
\begin{abstract}
We study how the charge neutrality affects the phase structure of
 three-flavor PNJL model. 
We point out that, within the conventional PNJL model at finite density
 the color neutrality is missing because the Wilson line serves as an
external ``colored'' field coupled to dynamical quarks. 
In this paper we heuristically assume that the model may still be
 applicable. 
To get color neutrality one has then to allow non vanishing color
 chemical potentials. 
We study how the quark matter phase diagram in $(T,m_s^2/\mu)$-plane is
 affected by imposing neutrality and by including the Polyakov loop
 dynamics.
Although these two effects are correlated in a nonlinear way, the impact
 of the Polyakov loop turns out to be significant in the $T$ direction,
 while imposing neutrality brings a remarkable effect in the $m_s^2/\mu$
 direction.
In particular, we find a novel unlocking transition, when the
 temperature is increased, even in the chiral $SU(3)$ limit.
We clarify how and why this is possible once the dynamics of the colored
 Polyakov loop is taken into account.
Also we succeed in giving an analytic expression for $T_c$ for the
transition from two-flavor pairing (2SC) to unpaired quark matter
 in the presence of the Polyakov loop.
\end{abstract}
\maketitle

\section{Introduction}
Quantum chromodynamics (QCD) is expected to exhibit a variety of phases
depending on the temperature and on the baryon density
\cite{reviews}. 
At zero density and finite temperature the two main features of QCD are
the confinement/deconfinement phase transition and chiral symmetry
restoration. 
They should be realized  when the hadronic system is heated, for example
in ultrarelativistic heavy ion scattering processes such as at the
Relativistic Heavy Ion Collision (RHIC) experiment \cite{RHIC} or in the
future ALICE experiment at the Large Hadron Collider (LHC) at CERN. 
Moreover this behavior is clearly seen by lattice QCD simulations
\cite{latticeTc}. 
In these conditions  quarks and gluons should be released as active
degrees of freedom at some critical temperature. 
Moreover in the same range of temperatures one expects the 
restoration of chiral symmetry, whose spontaneous breakdown is known 
to play a key role in the mass spectroscopy of zero density QCD
\cite{Nambu:1961tp,Hatsuda:1994pi}. 
At finite baryon densities similar transitions are also expected
although the comparison with lattice simulation data is still lacking
due to the so called fermion sign problem. 
However, at extremely high density, where 
perturbative techniques are allowed, it is now theoretically well
established that quarks are deconfined forming diquark condensates 
so that the system is in a color superconducting ground state with asymptotic 
color-flavor locking (CFL) \cite{Alford:1998mk}. 
While difficult to achieve in the laboratory color superconductivity
might be relevant to the inner structure of compact stellar objects
\cite{Weber:2004kj}.

Exploring phase structure at intermediate density, where neither 
lattice simulations nor perturbative calculations can be trusted, is 
the object of various model studies. 
There are different effective models which provide simple descriptions
of chiral symmetry restoration at finite temperature and density; 
the Nambu-Jona Lasinio (NJL) model is one of them
\cite{Nambu:1961tp,Hatsuda:1994pi,Asakawa:1989bq,Rapp:1997zu}. 
The NJL model realizes the spontaneous chiral symmetry breaking of QCD
at small temperature and density. 
Despite its simple structure, it can also realize a CFL phase at the
largest density. 
Moreover it can even reproduce the correct ratio of the gap and critical
temperature for the transition from the CFL to the unpaired phase.

The main defect of the NJL model is the absence of the
confinement/deconfinement transition. 
A theoretical attempt to understand the nature of the deconfinement
transition goes back to the work \cite{Polyakov:1978vu} in which
deconfinement was shown to be associated with the spontaneous breaking of the 
global $Z(N_c)$-symmetry of a finite temperature $SU(N_c)$ pure gauge
theory. 
The order parameter is the traced Polyakov loop, whose 
condensation and correlation are related to the free energy of 
static quark and the string tension between two static quarks in a
thermal medium.

The inclusion of the Polyakov loop dynamics into the NJL model was 
first done by Fukushima \cite{Fukushima:2003fw} in order to study 
the relation between chiral restoration and deconfinement. 
It is now called ``Polyakov loop extended Nambu-Jona Lasinio'' (PNJL) model. 
In this model, the chiral condensate $\bar{q}q$ serves as an order
parameter for the chiral transition, while the traced Polyakov loop 
$\Phi$ 
performs this job for the deconfinement transition. 
Even though the former and the latter have their definite meanings, as
order parameters, only within different limits, ($m_q\to 0$ and $m_q\to\infty$), 
they are still useful as indicators of both 
crossovers and/or transitions. 
In addition, the model enables to interpret nicely some bulk properties
of matter observed on the lattice on the field theoretical ground
\cite{Ratti:2005jh}.

The purpose of this work is to investigate the color superconducting
phase structure in $(T,m_s^2/\mu)$-plane within  the PNJL model, and 
to study the effect of imposing neutralities, in both the paired and
unpaired phases, in presence of the Polyakov loop. 
The neutrality constraints are known to be important for the candidates
of color superconducting phases at a realistic density; they open a
window in a phase diagram to intriguing gapless phases
\cite{Alford:2003fq,Shovkovy:2003uu}. 
Although a few works have already explored the pairing phases of PNJL
models \cite{Roessner:2006xn,Ciminale:2007ei,Ciminale:2007sr}, none of
them takes into account either the possibility of complicated gap
structures or the neutrality effects. 
Thus our study is a natural extension of them. 
One surprising result is that once PNJL model is applied to finite
density, it inevitably lacks color neutrality even when the system is
unpaired. 
This is a sort of sign problem at finite density. 
In this paper we still proceed on the heuristic assumption that PNJL is
applicable to finite densities. 
Also it will turn out that the inclusion of the Polyakov loop greatly
affects the phase diagram by stabilizing the two-flavor pairing (2SC)
phase, and it also brings about a color-flavor unlocking transition
\cite{Alford:1999pa} 
at finite temperature in a new mechanism.

The paper is organized as follows. 
In Sec.~\ref{form}, we introduce our model and approximations.
In the first part of Sec.~\ref{reso}, we demonstrate the lack of color
neutrality in the conventional PNJL model at finite density.
The rest of the section is devoted to discussion of the numerical results.
The neutrality effects, the effect of dynamics of Polyakov loop, and
their interplay will be particularly covered. 
We summarize the main contents of our paper with some concluding remarks in
Sec.~\ref{sum}.

\section{Formalism}\label{form}
In order to accommodate for pairing in the $J^P=0^+$ channel at 
finite density, we add the 4-point vertex to the free part of the
Polyakov-quark model, which hereafter we shall refer to as the 
Polyakov NJL (PNJL) model. 
\beq
  {\mathcal L}_{\rm eff}[q,\bar{q}; A_4]=\bar{q}(i(\ds{\mathcal D}[A_4]
  +\gamma_0(\mu+\delta\mu_{\rm eff}))q%
  +\frac{G}{4}\bar{q}P_\eta\bar{q}^Tq^T\bar{P}_\eta q%
  -{\mathcal U}(T,\Phi[A_4],\Phi[A_4]^*).
\eeq 
Here $q$ stands for the quark field, and a summation over color and
flavor degrees of freedom is understood.
$P_\eta=C\gamma_5\epsilon_{\eta ij}\epsilon_{\eta ab}$
($\bar{P}_\eta=\gamma_0 P_\eta^\dagger\gamma_0$) 
is the matrix, antisymmetric in color, flavor and spin, specifying the
pairing channel. 
The constant $G$ parameterizes the strength of the coupling leading to
diquark condensation. 
We work within the chiral $SU(2)$ limit, setting $m_u=m_d=0$, 
and take into account the strange quark mass within the high density
approximation. 
This means that we include the effect of its finite value in  the
chemical potential difference $\delta\mu_{\rm eff}$ \cite{Alford:2003fq}. 
As a result,
\beq
\ba{rcl}
  \delta\mu_{\rm eff}=%
  -\mu_eQ+\mu_3 T_3+\mu_8 T_8-\frac{m_s^2}{2\mu}\diag.(0,0,1)_f\times{\bf 1}_c,
\ea
\label{chem.pot} 
\eeq 
where $Q=\diag.(2/3,-1/3,-1/3)_f\times{\bf 1}_c$,
$T_3={\bf 1}_f\times\frac{1}{2}\lambda_3$, 
and $T_8={\bf 1}_f\times\frac{1}{\Sqrt{3}}\lambda_8$, 
with $\{\lambda_\alpha\}$ being
the standard Gell-Mann matrices. 
We find it more transparent switching to a new spinor basis for the
quark field defined as 
$q_A=(q_{ur},q_{dg},q_{sb},\,q_{ug},q_{dr},\,q_{sr},q_{ub},\,q_{db},q_{sg})$
by means of:
\beq
q_{i\alpha}=\sum_{A=1}^9(F_A)_{i\alpha}q_A
\eeq
with $(F_A)_{i\alpha}$ unitary matrices in color and flavor space
defined in \cite{Casalbuoni:2004tb}.
In this new basis \eqref{chem.pot} takes the form of a diagonal matrix:
$\delta\mu^A_{\rm eff}\delta_{AB}$
with $A=1(ur),\,2(dg),\,\cdots,\,9(sg)$.

We treat the Polyakov loop by the static, homogeneous, and 
classical background gauge field
$A_4\equiv ig{\mathcal A}_0^\alpha\frac{\lambda_\alpha}{2}$
where the temporal gauge field $A_4$ is introduced by parameterizing the
Wilson-line, as $L=e^{iA_4/T}$. 
In the PNJL model one assumes that this background gauge field couples
to quarks with covariant derivative
${\mathcal D}_\mu=\partial_\mu-\delta_{\mu 0}A_4$. 
In the convenient gauge called Polyakov gauge, the Wilson line $L$ is in
the diagonal representation \cite{Fukushima:2003fw},\footnote{%
One can always find the gauge rotation $U$ such that $ULU^{-1}$
becomes diagonal.
This is simply a gauge fixing, and any physical quantities will
not depend on the gauge freedom $U$ so we can safely reduce the eight
dynamical variables to parametrize $L$ up to two independent parameters 
$\{\phi_3,\phi_8\}$.
We note, however, when the diquark condensation is taken into account,
this is no longer justified unless more general ansatz for the
diquark condensation, $\Delta_{\eta\eta'}\epsilon_{\eta ab}%
\epsilon_{\eta'ij}$, is adopted.
Simultaneous color-flavor rotation can make the condensate matrix
diagonal such that the usual assumption, 
$\Delta_{\eta\eta'}\propto\delta_{\eta\eta'}$ ,is recovered, but this is
nothing but the gauge fixing.
Thus in principle, if we adopt the diagonal ansatz for the diquark
condensation, we can no longer make $L$ gauge rotated to diagonal, and
on the other hand, if we adopt the diagonal form of $L$, we should work in
more general assumption for $\Delta_{\eta\eta'}$.
Nevertheless, we work in the simplified assumption that both $L$ and
$\Delta_{\eta\eta'}$ are of diagonal as in \cite{Roessner:2006xn}
leaving a further detailed analysis in the future.} 
\ie,
\beq
 L=e^{(i\phi_3\lambda_3+i\phi_8\lambda_8)/T}.
\label{eq:diagonal}
\eeq
Moreover we restrict ourselves to the case $\phi_8=0$ such that
the traced Polyakov loop $\Phi=\tr_c L/N_c$ becomes real
\cite{Roessner:2006xn} 
whereas at finite density there is no strict reason why $\Phi$ should be
real \cite{Dumitru:2005ng,Fukushima:2006uv}.
Thus in this representation, $\Phi=\frac{2\cos(\phi_3/T)+1}{3}$, and
the effect of the background field $A_4$ is just to shift the color
chemical potential to the imaginary direction
$\mu_3\to\mu_3-2i\phi_3\equiv\tilde{\mu}_3$. 
For the Polyakov loop effective potential ${\mathcal U}$ we use the
following form, inspired by the strong coupling analysis of the pure gauge 
sector \cite{Polonyi:1982wz,Fukushima:2003fw,Ghosh:2007wy}
\beq
\frac{{\mathcal U}(T,\Phi,\Phi^*)}{T^4}=-\frac{b_2(T)}{2}\Phi^*\Phi%
  +b(T)\log\left(1-6\Phi^*\Phi+4(\Phi^{*3}+\Phi^3)-(\Phi^*\Phi)^2\right),
\eeq
with
\beq
  b_2(T)=a_0+a_1\left(\frac{T_0}{T}\right)+a_2\left(\frac{T_0}{T}\right)^2,
 \quad b(T)=b_3\left(\frac{T_0}{T}\right)^3.
\eeq
The numerical values for coefficients are determined by fitting several
quantities to the lattice results of pure gauge theory \cite{Ratti:2005jh}:
\beq
 a_0=3.51,\quad a_1=-2.47,\quad a_2=15.2,\quad, b_3=-1.75.
\eeq 
In the absence of dynamical quarks, $T_0$ is set to the value of the
transition temperature for deconfinement, \ie, $T_0=270\MeV$. 
In our model, we use the value $T_0=208\MeV$ which is the theoretically
suggested value for $T_0$ in the presence of two light flavors, $N_f=2$
\cite{Ratti:2005jh,Schaefer:2007pw}, 
although as we treat the strange quark mass as a free parameter, our
calculation will cover the situations between two flavor and three
flavor, \ie, $N_f=2\,(+1)\to 3$.
In the case $N_f=3$, the slightly lowered value $T_0=178\MeV$ is
proposed, but we have checked that choosing this value for $T_0$ does
not change our results in any significant way.

By introducing a charge conjugated field $q_c=-Cq^*$ as an 
independent field, and after introducing the Hubbard-Stratonovich
field $\Delta_\eta(\tau,\bfm{x})=\frac{G}{2}q^T\bar{P}_\eta q$ and
$\bar{\Delta}_\eta(\tau,\bfm{x})=\frac{G}{2}\bar{q}P_\eta\bar{q}^T$,
we integrate out the fermion field $Q=(q,q_c)$. 
Within the mean field approximation for $\Delta$ and $\bar{\Delta}$, the
effective potential becomes
\beq
\ba{rcl}
 \Omega(\Delta_\eta,A_4,\mu_e,\mu_3,\mu_8;\mu,T)%
 &=&{\mathcal U}(T,\Phi,\Phi^*)%
  -\frac{\mu_e^4}{12\pi^2}-\frac{\mu_e^2T^2}{6}-\frac{7\pi^2 T^4}{180}\\[2ex]
 & &+\sum_\eta\frac{\Delta_\eta^2}{G}%
 -\frac{1}{2}\ln\Det\left(\begin{array}{cc}
  i\gamma_0\ds{D}[A_4]+{\mu}+\delta\mu_{\rm eff}&%
  -\Delta_\eta\gamma_5\epsilon_{\eta ij}\epsilon_{\eta ab}\cr
  -\Delta_\eta\gamma_5\epsilon_{\eta ab}\epsilon_{\eta ij}&
  i\ds{D}[-A_4]^t\gamma_0-{\mu}^t-{\delta\mu}_{\rm eff}^t\cr
 \end{array}\right).
\ea 
\eeq
where
$D_\mu[A_4]={\mathcal D}_{\mu}[A_4]|_{\partial_0\to i\partial_\tau}$
with $\tau$ denoting the imaginary time, and the transpose operation $X^t$ 
only acts on the color and flavor structure of $X$.

It is useful to write down the thermodynamic potential in the
$\Delta_\eta=0$ 
and $\mu_{e,3,8}=0$ case. 
Leaving the trace over color, it takes a form
\beq
 \Omega={\mathcal U}(T,\Phi,\Phi^*)
-2N_fT\int\frac{d\bfm{p}}{(2\pi)^3}\tr_c{\ln}%
 \left[(1+L^\dagger e^{-(p-\mu)/T})
 (1+L e^{-(p+\mu)/T})\right].
\label{eq:bf}
\eeq 
with $N_f=3$. 
Within the Polyakov gauge and imposing the $\phi_8=0$ prescription, the
Wilson line takes the following form
\beq
 L=\left(%
 \begin{array}{ccc}
 l& &\cr
 &l^*&\cr
 &&1\cr
\end{array}\right),
\label{Ldiag}
\eeq 
with $l=e^{i\phi_3/T}$.

Let us now consider the case with $\Delta_\eta\neq 0$ and
$\mu_{e,3,8}\neq0$. 
We try to simplify the expression for the thermodynamic potential. 
Within the current approximation (treating $m_s\neq 0$ as a shift of the
chemical potential), the action does not mix the left-handed quark and
the right handed quarks. 
Thus, we can rewrite the functional determinant as 
\beq
 -\frac{1}{2}\ln\Det\left(\begin{array}{cc}
  i\gamma_0\ds{D}[A_4]+{\mu}+{\delta\mu}_{\rm eff}&%
  -\Delta_\eta\epsilon_{\eta ij}\epsilon_{\eta ab}\cr
  -\Delta_\eta\epsilon_{\eta ab}\epsilon_{\eta ij}&
  i\ds{D}[-A_4]^t\gamma_0-{\mu}^t-{\delta\mu}_{\rm eff}^t\cr
 \end{array}\right)_L - (L\to R,\,\Delta_\eta\to-\Delta_\eta).
\eeq
Now the Dirac gamma matrices can be regarded as two dimensional matrices
for Weyl spinors, say, $\gamma_\mu=(\bf{1},\bfm{\sigma})$. 
Finally, putting $\bfm{p}=\mu \bfm{v}+\bfm{l}$ with velocity
$|\bfm{v}|=1$, 
and discarding the antiquark contributions, 
we get the high density effective theory (HDET) approximation for the
effective potential \cite{Nardulli:2002ma}.
\beq
\ba{rcl}
 \Omega(\Delta_\eta,A_4,\mu_e,\mu_3,\mu_8;\mu,T)%
 &=&{\mathcal U}(T,\Phi,\Phi^*)%
  -\frac{\mu_e^4}{12\pi^2}-\frac{\mu_e^2T^2}{6}-\frac{7\pi^2 T^4}{180}\\[2ex]
 & &+\sum_\eta\frac{\Delta_\eta^2}{G}%
 -\frac{T}{2}\sum_n%
  \int\frac{d\bfsm{v}}{4\pi}%
  \int_{-\omega_c}^{\omega_c}\frac{\mu^2dl_{\parallel}}{2\pi^2}%
  \ln S^{-1}_{L,+}(i\omega_n,\bfm{v}\cdot\bfm{l})-(L\to R).
\ea 
\eeq 
where $\omega_c$ is a momentum cutoff, and the positive energy left
handed projected propagator is defined as
\beq
 S^{-1}_{L,+}(i\omega_n,\bfm{v}\cdot\bfm{l})%
 =\left(\begin{array}{cc}
   i\omega_n-\bfm{v}\cdot\bfm{l}+\delta\mu_{\rm
    eff}-iA_4&-\Delta_\eta\epsilon_{\eta ab}\epsilon_{\eta ij}\cr
   -\Delta_\eta\epsilon_{\eta ab}\epsilon_{\eta ij}%
   &i\omega_n+\bfm{v}\cdot\bfm{l}-\delta\mu_{\rm eff}^t+iA_4^t\cr
   \end{array}
  \right).
\label{eq:Sinv}
\eeq 
This is now a $18\times 18$ matrix defined in the color-flavor space.
It has a form $(i\omega_n{\bf 1}_{18}-{\mathcal H})$.
In order to evaluate the Matsubara summation, we have to evaluate all
the eigenvalues of the hamiltonian density ${\mathcal H}$. 
Since we have doubled the degrees of freedom by introducing the
Nambu-Gorkov notation, the eigenvalues of the hamiltonian will appear as
$\{E_A(l_{\parallel}),-E_A(l_{\parallel})\}$
with $A=1,2,\cdots,9$. 
In contrast to the standard NJL models without Polyakov loop, 
${\mathcal H}$ 
is no longer Hermitian due to the imaginary chemical 
potential $\tilde{\mu}_3$, and accordingly each quasiparticle energy
$E_A(l_\parallel)$ 
can take in general complex values. 
Consequently, $\Omega$ is no longer restricted to be real. 
We avoid this sign problem by taking the real part of $\Omega$ as in
\cite{Ratti:2005jh}. 
Once these quasiparticle energies are evaluated, and choosing the basis
as $\{E_A,-E_A\}$ such that both the conditions, $\Re E_A\ge 0$
and $E_A\to|l_{\parallel}-\delta\mu_{\rm eff}^{A}|$
when $\Delta_\eta,\phi_3\to 0$, are satisfied, 
we can perform the Matsubara summation as 
\beq
\ba{rcl}
 \Re\Omega(\Delta_\eta,A_4,\mu_e,\mu_3,\mu_8;\mu,T)%
 &=&{\mathcal U}(T,\Phi,\Phi^*)%
  -\frac{\mu_e^4}{12\pi^2}-\frac{\mu_e^2T^2}{6}-\frac{7\pi^2 T^4}{180}%
  -\sum_{A=1}^{9}\frac{(\mu+\delta\mu_{\rm eff}^{A})^4}%
  {12\pi^2}\\[2ex]
 & &+\sum_\eta\frac{\Delta_\eta^2}{G}%
 -\sum_{A=1}^9%%
  \int_{-\omega_c}^{\omega_c}\frac{\mu^2dl_{\parallel}}{2\pi^2}%
  \Big[\Re E_A(l_\parallel)%
  -|l_{\parallel}-\delta\mu_{\rm eff}^{A}|%
  +2T\ln(|\!|1+e^{-E_A(l_\parallel)/T}|\!|)\Big].\\[2ex]
\ea 
\eeq
We took the energy density in the vacuum without any condensation as the
reference energy density. 
The fifth term is nothing but the zero temperature part of free quark
contribution to the effective potential. 
We need the ultra violet cutoff $\omega_c$ 
only in the integral representing the condensation energy, \ie, the last
term.\footnote{The integral of the thermal part ($T\ln(\cdots)$) can be
evaluated without cutoff.}
Instead of $G$, we use $\Delta_0$, the zero temperature CFL gap in the
chiral $SU(3)$ limit without Polyakov loop, as the indicator of diquark
attraction \cite{Schafer:1999fe}
\beq
 \frac{1}{G}=\frac{2\mu^2}{\pi^2}%
 \ln\left(\frac{2\omega_c}{2^{1/3}\Delta_0}\right).
\label{mcoupling}
\eeq
With the use of this cutoff dependent coupling constant, the derivatives
of the effective potentials,
$\frac{\partial\Re\Omega}{\partial(\Delta_\eta,\phi_3)}$,
now have well-defined limits as $\omega_c\to\infty$.
In this way, we can remove the cut-off dependence from the gap
equations, while it remains in the effective potential itself.

The evaluation of the effective potential is carried out by finding the
eigenvalues of ${\mathcal H}$ for given momentum $l_{\parallel}$,
and integrating them over the momentum. 
Then the mean field solution for the ground state is obtained by
minimizing the effective potential with respect to
$(\Delta_\eta,\phi_3)$
imposing the proper constraints of charge neutrality,
$\frac{\partial\Re\Omega}{\partial(\mu_e,\mu_3,\mu_8)}=0$.

\section{Results}\label{reso}
In this section, for the numerical computations we fix $\mu=500\MeV$,
and concentrate on the case with $\Delta_0=60\MeV$ with $\omega_c=300\MeV$. 
We would study the phase diagram, and how the physical quantities behave
as functions of $\frac{m_s^2}{2\mu}$, and temperature $T$, treated as
free parameters. 
In the numerical calculations, in order to take the minimum strong
coupling effect into account, we bring back the momentum dependence of
the density of state
$\frac{\mu^2}{2\pi^2}\to\frac{(l_{\parallel}+\mu)^2}{2\pi^2}$. 
By doing this, the particle-hole asymmetry which is known as the first
correction to the weak coupling approximation will be properly taken
into account.
 \subsection{Color neutrality}\label{resoneutrality}
Before discussing in detail the calculation let us briefly comment on a
strange feature that shows up, even in the unpaired phase ($\Delta_\eta=0$), 
if one assumes vanishing charge chemical potentials. 
The interesting fact is that the conventional PNJL model calculation at
finite $\mu$ lacks color neutrality. 
It is worth stressing here that this does not follow from HDET
approximation (see \eqref{eq:bf} obtained before introducing HDET
formalism).
It is clear from \eqref{Ldiag} that the Wilson line couples to each
color of quarks, $(r,g,b)$, with different weight. 
Looking at the real part only, the weight for $(r,g)$ quarks differs
from that of $b$ quarks. 
Thus, the energy required to populate $(r,g)$ quarks in the background
gauge field $A_4$ is different from that for $b$ quarks. 
Furthermore, $\cos(\phi_3/T)$ can take a negative value close to $-1/2$
when the system is nearly confined. 
This means that thermal excitations of on-shell $(r,g)$ quarks reduce the 
pressure of the system making it difficult to create on-shell $(r,g)$ quarks 
in the system.  
This can be regarded as the effect of confinement in this PNJL model at
finite density. 
The problem is that, even in this situation, $b$ quarks can be excited
at finite temperature because $L_{33}$ is unity and thus the thermal
weight for $b$ quarks does not differ from that in the deconfined phase.
This unphysical feature might be viewed as an artifact of this PNJL
model which originates in the assumption that the dynamics of traced
Polyakov loop $\Phi$ can be equivalently described by the constant
background gauge field $A_4$. 
As a consequence, the development of the finite value of $A_4$
breaks not only the $Z(3)$ center symmetry but also the color $SU(3)$
symmetry in a spontaneous way; this fact does not contradict the
Elitzur's theorem \cite{Elitzur} since we still expect physical
quantities such as quasi-particle dispersions should not depend on the
gauge and thus have their definite meanings even after gauge unfixing
which would make $A_4$ itself vanish.

\begin{figure}[tp]
  \includegraphics[width=0.457\textwidth,clip]{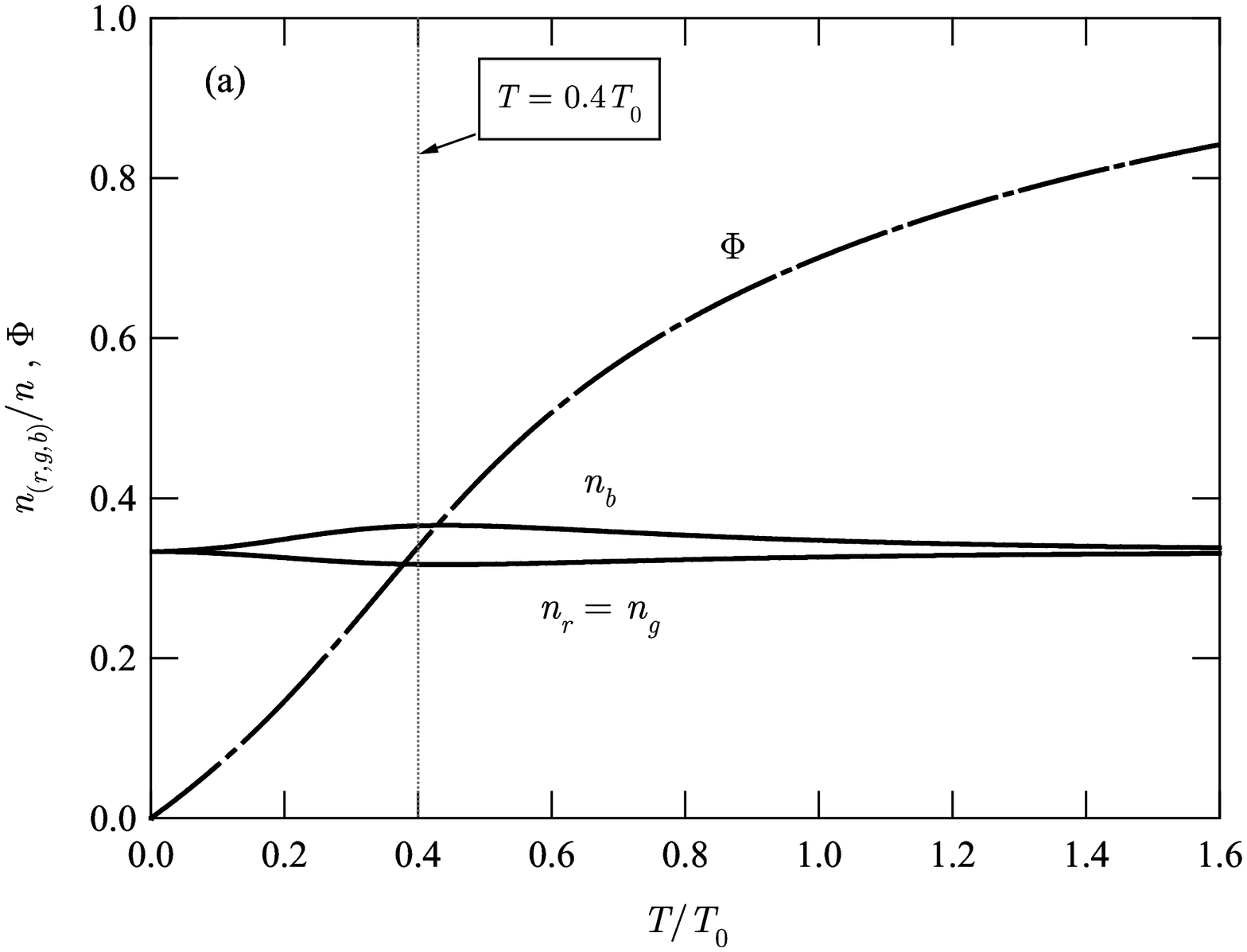}
  \includegraphics[width=0.457\textwidth,clip]{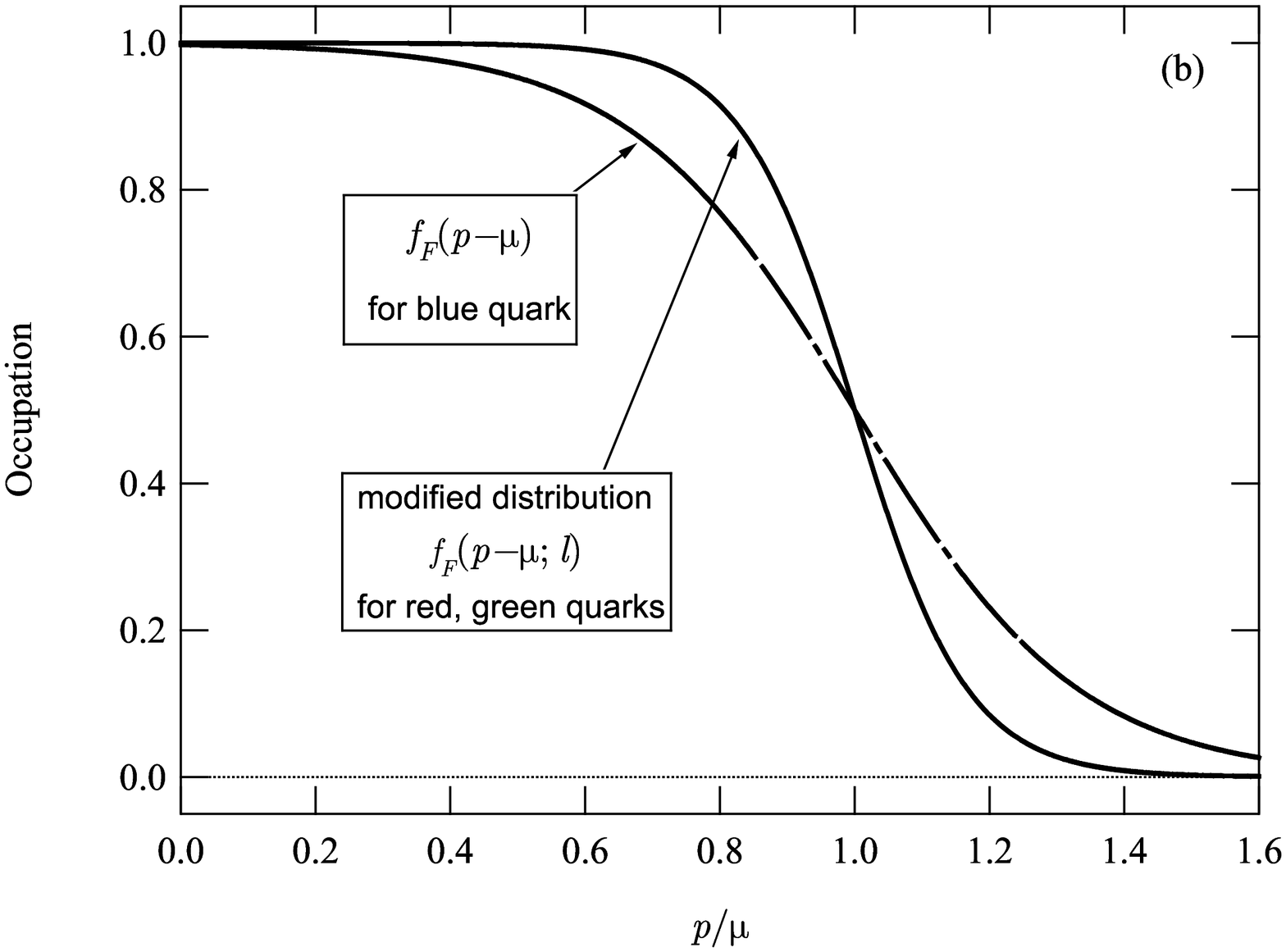}
 \caption[]{(a):~Each color density of a finite density PNJL model
 with $m_s=\Delta_\eta=\mu_{e,3,8}=0$ as a function of $T$. $T_8$-color
 density is induced by the antitriplet color charge $l^*$
 of Wilson line coupled to quarks.
 (b):~The occupation number profile of quark with red, green, and blue,
 as a function of momentum at $T=0.4T_0$ which is indicated by the
 vertical dotted line in (a).}
 \label{colordensity}
\end{figure}

One may think that this undesired feature is just due to the wrong
ansatz of the PNJL model itself.
Apart from such a possibility, in order to avoid an unphysical
appearance of color density within the model, we should inevitably
introduce an appropriate color chemical potential. 
According to the gauge of the Polyakov loop, in our case, $\mu_8$ is
required to maintain the unpaired phase color neutral. 
To be more explicit, ignoring the antiquarks we can write the color
density as 
\beq 
\ba{rcl}
  n_{r,g}&=&2N_f\int\frac{d\bfsm{p}}{(2\pi)^3}%
  f_F(p-\mu;l)-(l\to
  l^{*},\mu\to-\mu)\\[2ex]
  &=&n_b+2N_f\int\frac{d\bfsm{p}}{(2\pi)^3}%
  \frac{3(\Phi-1)\tanh\left(\frac{p-\mu}{2T}\right)}%
  {6\Phi+4\cosh\left(\frac{p-\mu}{T}\right)}-(\mu\to-\mu),
\ea 
\label{integ} 
\eeq 
where $n_b=2N_f\int\frac{d\bfsm{p}}{(2\pi)^3}\frac{1}{e^{(p-\mu)/T}+1}$
is just the density of a free Fermi gas and
\beq
f_F(p-\mu;l)\equiv\Re\big(\frac{1}{l^{-1}e^{(p-\mu)/T}+1}\big)
\label{modifiedfermi}
\eeq 
is the modified Fermi distribution in the presence of $A_4$ 
describing $(r,g)$ quarks which is plotted in Fig.~\ref{colordensity}(b)
compared to the standard Fermi distribution $f_F(p-\mu)$ for blue quarks.
From \eqref{modifiedfermi} it follows that the density of $r$ and $g$
quarks differs from that of simple fermi gas, and the difference never
disappears unless either $T\to0$ or $\Phi\to1\,(T\to\infty)$ is approached. 
The difference also cancels at $\mu=0$ thanks to the equal and opposite
contribution from antiquarks.

In Fig.~\ref{colordensity}(a), we plot the ratios of each color
density, $n_r$, $n_g$, and $n_b$, to the total quark density as a
function of $T$ in the unpaired phase with $\mu_{e,3,8}=0$. 
In the numerical evaluation we set $\mu=500\MeV$, and ignore the tiny
antiquark contributions. 
As discussed above, we can see the finite difference between $n_r=n_g$
and $n_b$ in the intermediate range of temperature.\footnote{%
It should be noted that the color density itself is a gauge
dependent quantity and thus should depend on the choice of the
gauge parametrizing the Wilson line. 
With our diagonal representation of \Eqn{eq:diagonal} with $\phi_8=0$,
the $T_8$ color density becomes finite as we observed above.
If we selected the different gauge, the other entry of octet color
densities $\{\langle q^\dagger T_\alpha q\rangle\}$ should have appeared.
The important thing is, however, whichever gauge we
choose,some color density should become finite; in fact the
squared sum of the octet color densities is shown to be the gauge
independent quantity \cite{Buballa:2005bv}.}
The deviation becomes maximum near the steepest point of the Polyakov
loop, which is usually identified as the deconfinement transition
\cite{Ratti:2005jh}.

\subsection{Phase diagrams}
\begin{figure}[tp]
  \includegraphics[width=0.457\textwidth,clip]{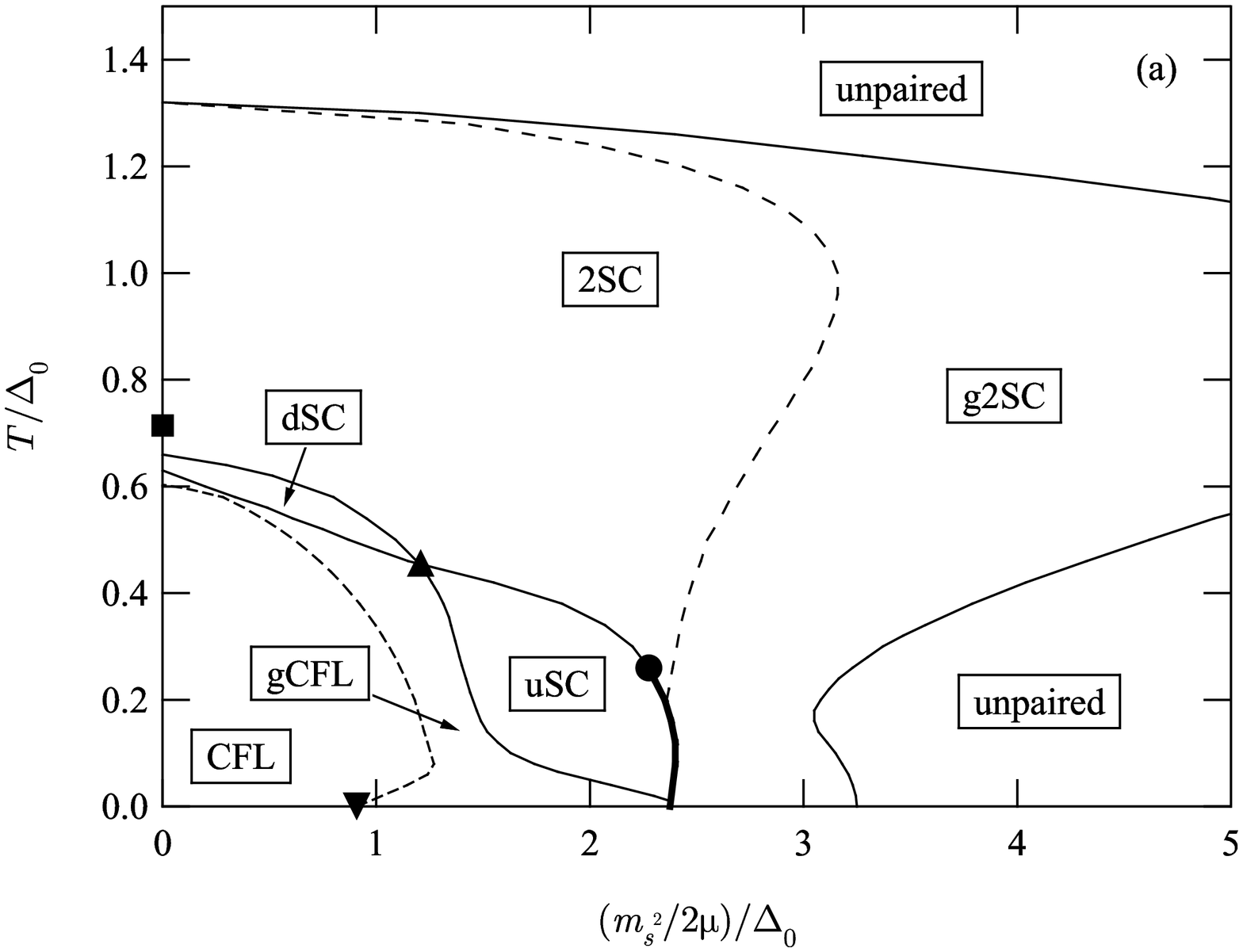}
  \includegraphics[width=0.457\textwidth,clip]{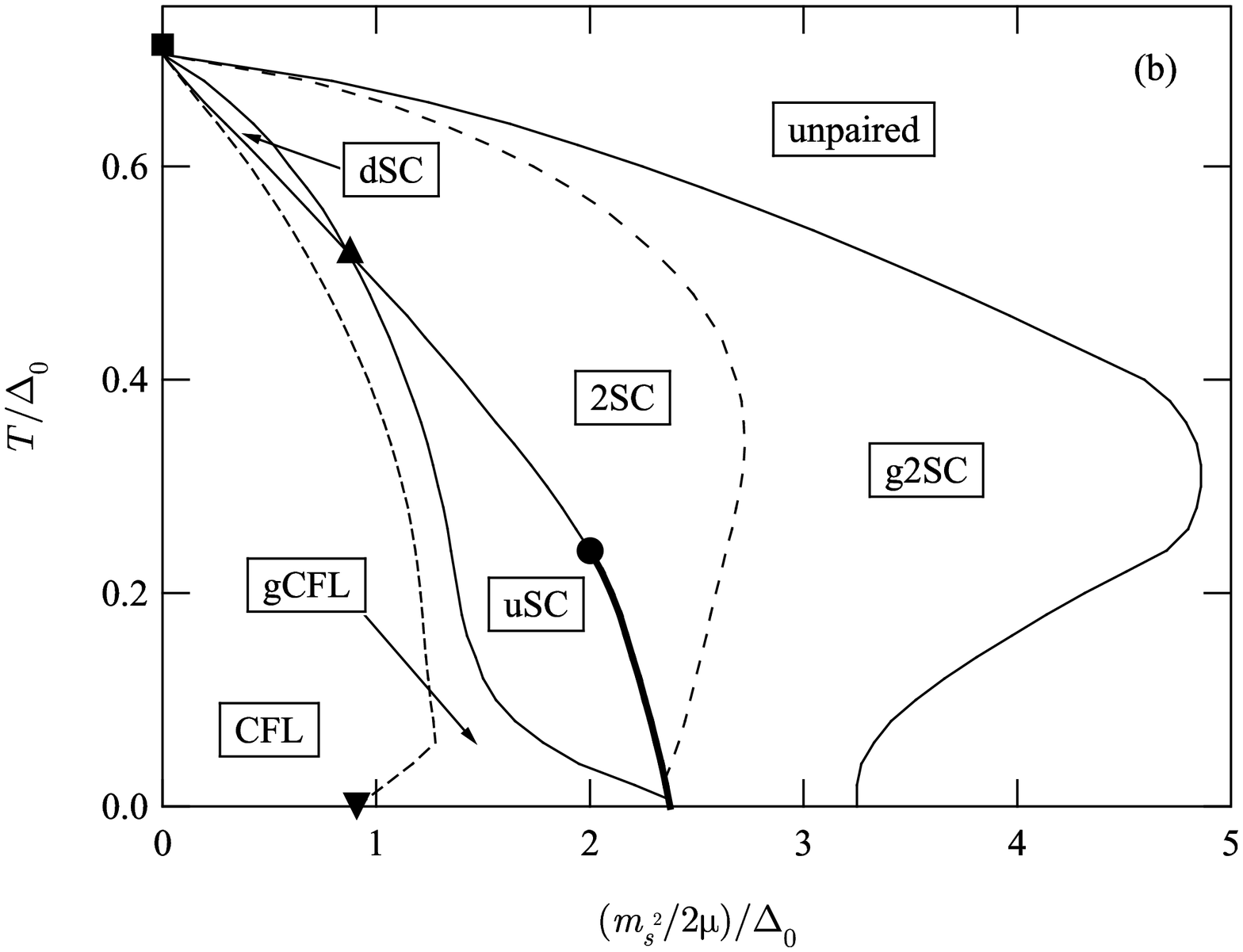}
 \caption[]{(a):~Phase diagram in $(\frac{m_s^2}{2\mu},T)$-plane
 at $\Delta_0=60\MeV,\,\mu=500\MeV$ with the Polyakov loop under charge
 neutrality constraints. 
 The energy scales are normalized by $\Delta_0$.
 The bold line corresponds to 1st order phase transition while the thin line 
 indicates the 2nd order phase transition.
 For other instructions, see text.
 (b):~The same as (a) but without the Polyakov loop.
 In both figures, the bold square put on the $T$ axis indicates the weak
 coupling approximation of the critical temperature at chiral limit
 without $\Phi$, \ie, $T_c^0/\Delta_0=0.714$. }
 \label{phase}
\end{figure}

In Fig.~\ref{phase}(a), the phase diagram in
the $\left(T,\frac{m_s^2}{2\mu}\right)$-plane of the charge neutral PNJL
model is displayed. 
The bold line represents the first order phase transition, while the thin 
line represents the second order phase transition. 
The dashed line corresponds to the crossover from the 2SC phase to the
gapless 2SC (g2SC) phase. 
For comparison, we have shown in figure \ref{phase}(b) the phase diagram 
for the charge neutral NJL model which is the same reported in
\cite{Fukushima:2004zq}. 
We are specifying each phase as summarized in TABLE~\ref{phases}.
\begin{table*}[bp]
 \begin{tabular}{|r||c|}
  \hline
  \multicolumn{1}{|c||}{\bf\sl (g)CFL} &$\mathstrut$
  {$\Delta_\eta\ne 0$ for $\eta=1,2,3$} $\big($with
  $\Delta_2\le\big|\frac{\mu_e}{2}+\frac{\mu_3}{4}+\frac{\mu_8}{2}%
  -\frac{m_s^2}{4\mu}\big|$ or
  $\Delta_1\le\big|-\frac{\mu_3}{4}+\frac{\mu_8}{2}-\frac{m_s^2}{4\mu}\big|$
  satisfied$\big)$\\ \hline
  \multicolumn{1}{|c||}{\bf\sl dSC} &$\mathstrut$
  {$\Delta_{1,3}\ne0$, $\Delta_2^{\mathstrut}=0$}
  \\ \hline
  \multicolumn{1}{|c||}{\bf\sl uSC} &$\mathstrut$
  {$\Delta_{2,3}\ne0$, $\Delta_1^{\mathstrut}=0$}
  \\ \hline
  \multicolumn{1}{|c||}{\bf\sl (g)2SC} &$\mathstrut$
  {$\Delta_3\ne0$,
  $\Delta_1=\Delta_2=0$ $\big($%
  with
  $\Delta_3\le\big|\frac{|\mu_e|^{\mathstrut}-|\mu_3|^{\mathstrut}}{2}\big|$
  satisfied$\big)$}
  \\ \hline
 \end{tabular}
 \caption[]{The definition of pairing phases of current interest.
 }
 \label{phases}
\end{table*}
From these graphs, the impact of the Polyakov loop dynamics on the
pairing is quite obvious. 
The inclusion of the temporal gluon field significantly broadens the
region for the superconducting phase, in particular for the 2SC phase. 
In fact, the critical temperature is almost twice as large as
that in the NJL model without the Polyakov loop which is already known
in \cite{Roessner:2006xn}. 
(Note that the scale for the $T$-axis of figure (a) is twice as that of
figure (b).) 
Apart from this significant quantitative change, the qualitative
behavior of the phase diagram is not so much affected. 
In both cases, the d-quark superconducting phase ($\Delta_2=0$; dSC)
exists in a small region at finite temperature
\cite{Iida:2003cc,Fukushima:2004zq}, and there is the 
doubly critical point indicated by the upper triangle, a point
where the line for the vanishing of $\Delta_1$ intersects that 
for $\Delta_2$ \cite{Fukushima:2004zq}. 
Also the existence and the location of the critical point where the
fully gapped CFL phase turns into the gapless CFL (gCFL) phase phase is
not affected. 
In both figures the point is indicated by a lower triangle on 
the $\frac{m_s^2}{2\mu}$ axis. 
This fact means the the effect of the Polyakov loop on the pairing is
absent at $T=0$. 
The reason is that at $T=0$ the Polyakov dynamics decouples from the
pairing (NJL) dynamics so that the effect is absent because the temporal
gauge field $\phi_3$ is proportional to $T$ itself.

\subsection{Impact of the Polyakov loop at finite temperature;
  color-flavor unlocking and stiff 2SC phase}
\begin{figure}[tp]
  \includegraphics[width=0.457\textwidth,clip]{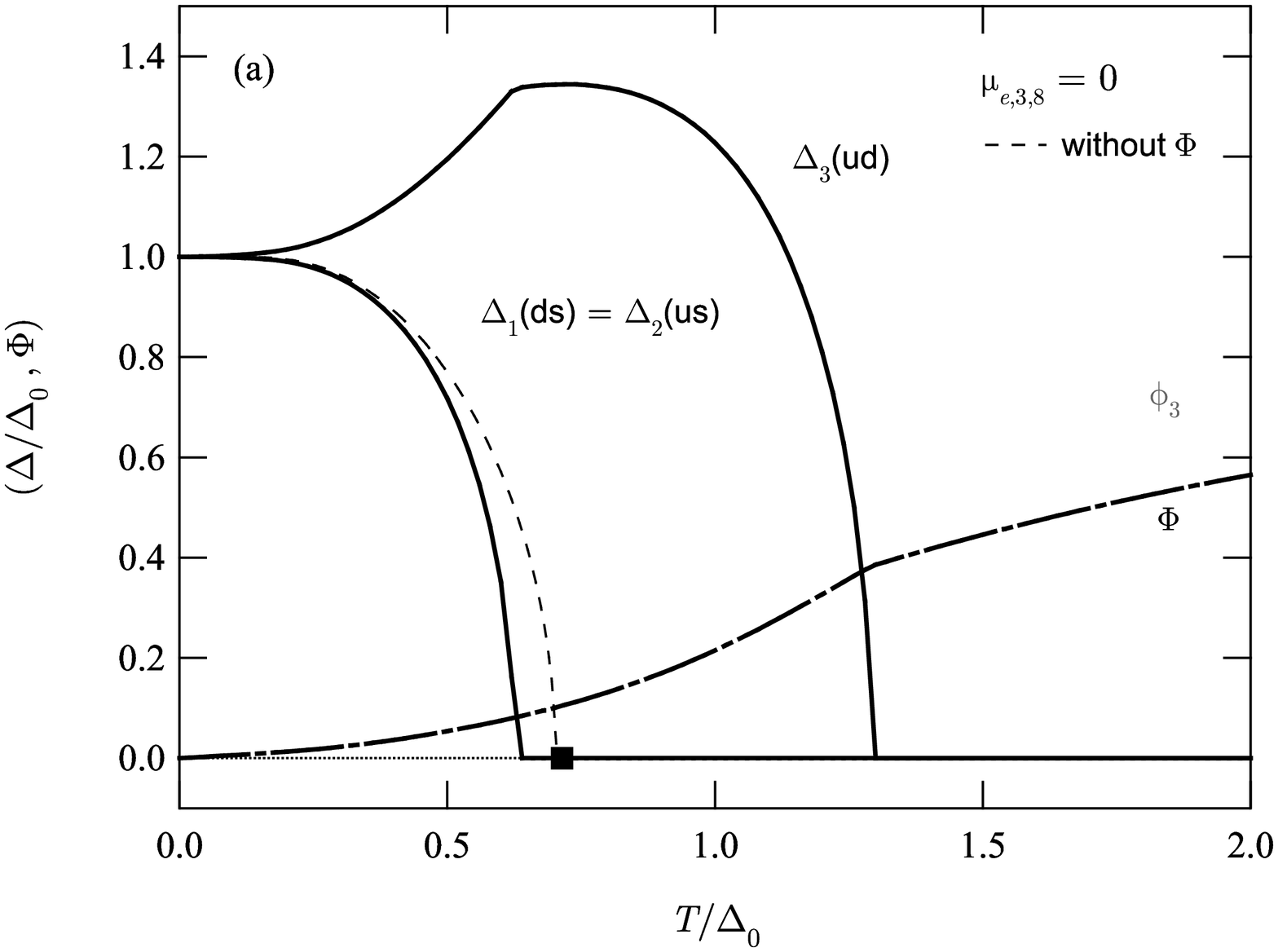}
  \includegraphics[width=0.457\textwidth,clip]{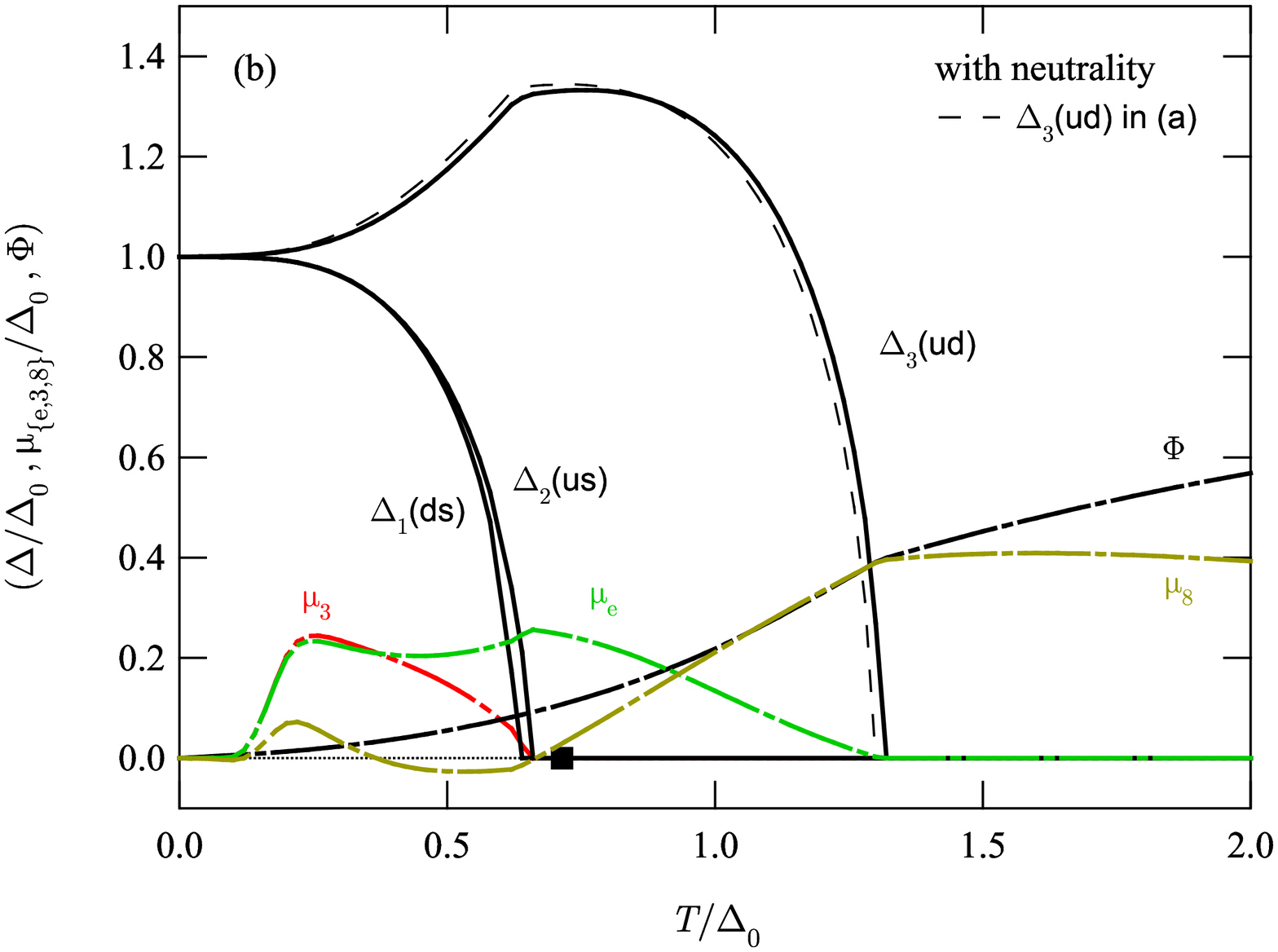}
 \caption[]{(a):~Gaps (solid lines) and the Polyakov loop (long dashed
  line) as functions of $T$ without the neutrality constraints, \ie,
  $\mu_{e,3,8}=0$.
  For the dashed line and the bold square, see the text.
 (b):~The same as (a) but with the neutrality constraints being
 respected. 
 Three charge chemical potentials are also depicted.
 $\Delta_3(ud)$
 in (a) is also shown by the dashed line, just for comparison.
 }
 \label{gapvsT}
\end{figure}

Next we focus in detail on the phase transitions at $m_s=0$ in order to
study the impact of the Polyakov loop dynamics and charge neutrality. 
To this end, we examine each effect step by step. 
In Fig.~\ref{gapvsT}(a), we show the gaps $\Delta_\eta(T)$ (solid lines)
and the Polyakov loop $\Phi(T)$ (long-dashed one) calculated without the
neutrality. 
For comparison, we also show by the dashed line the $\Delta_\eta(T)$
calculated with the NJL model without the Polyakov loop. 
In this case, the three gaps have the same behavior as functions of $T$
and they drop to zero simultaneously when $T_c^0\sim 0.714\Delta_0$
(shown by the bold square on the $\frac{m_s^2}{2\mu}$ axis) is approached. 
Once the Polyakov loop is taken into account this is no longer true as
one can see from the figure. 
Two gaps have the same magnitude, $\Delta_1(ds)=\Delta_2(su)$, 
while $\Delta_3(ud)$ is larger; moreover the two gaps $\Delta_1=\Delta_2$ 
drop to zero simultaneously near a  point lower than the bold
square ($T_c^0\sim 0.714\Delta_0$). 
This can be described as a second order color-flavor unlocking
transition induced by the Polyakov loop dynamics. 
This behavior is not strange because even though the Polyakov loop is
blind to the flavor degrees of freedom, in the color-flavor locked
phase, however, color is locked to flavor which is the way the Polyakov
loop affects the gap structure. 
In fact, as we have already discussed in Sec.~\ref{resoneutrality},
the presence of the  Polyakov loop induces a finite color $T_8$ density
as $n_r=n_g\ne n_b$ in the unpaired phase. 
This means that the existence of the Polyakov loop adds to the real part of 
the effective potential, $\Re\Omega$, the finite external field 
with $T_8$ charge, which explicitly breaks global color $SU(3)_c$ down
to $SU(2)_c$. 
Since color and flavor are locked in the CFL phase, the external field
induced by $\Phi$ tends to break the $SU(3)_{c+V}$ symmetry down to
$SU(2)_{c+V}$. 
This is the very reason for the splitting, $\Delta_1=\Delta_2\ne\Delta_3$ 
and also for the color-flavor unlocking to the 2SC phase. 
The emergence of color-flavor unlocking is one of the most interesting
features of the inclusion of the Polyakov loop dynamics in NJL model.

A further aspect deserves to be stressed. 
The 2SC phase persists up to $1.3\Delta_0$ which is almost twice as
large as the weak coupling formula for the critical temperature,
$T_c^0=0.714\Delta_0$ 
(indicated by the bold square). 
This striking feature is already noticed in the limiting case with the
pure CFL ansatz using the Ginzburg-Landau approach \cite{Ciminale:2007ei}. 
In what follows, we try to explain this fact. 
In the case of the 2SC case, we can explicitly determine the
quasiparticle energy dispersion. 
Four out of nine quasiquarks have nontrivial dispersion laws:
$E_{1,2}=\mp i\phi_3+\Sqrt{l_{\parallel}^2+\Delta_3^2}$
and $E_{4,5}=\pm i\phi_3+\Sqrt{l_{\parallel}^2+\Delta_3^2}$.
In this case, there is no shift of the averaged chemical potential to the 
imaginary direction. 
The gap equation in the absence of charge chemical potentials then
becomes
\beq
\frac{\pi^2}{2\mu^2}\left(\frac{1}{G}\right)=\int_{0}^{\omega_c}%
 dl_{\parallel}%
 \frac{1-2f_F%
 \Big[1-\frac{\frac{3}{2}(1-\Phi)(1-2f_F)(1-f_F)}{1-3(1-\Phi)f_F(1-f_F)}\Big]}
 {\Sqrt{l_{\parallel}^2+\Delta_3^2}},
\label{crit} 
\eeq 
where $f_F(x)=\frac{1}{1+e^{x/T}}$ is the Fermi distribution, and its
energy argument is now supposed to be $\Sqrt{l_\parallel^2+\Delta_3^2}$. 
The second term proportional to $f_F$ is referred to as a blocking integral 
due to thermally excited quarks. 
We need no cutoff $\omega_c$ for this part to be evaluated. 
We notice that the effect of the temporal gauge field is just to
suppress the blocking integral. 
This explains the robustness of 2SC in the presence of $\phi_3$, and 
the increase of the critical temperature as follows. 
We put $\Delta_3=0$ in \eqref{crit} and try to solve it in $T$ to 
derive the critical temperature. 
\eqref{crit} with $\Delta_3=0$ is nothing but the condition for
criticality, the Thouless criterion, which guarantees the divergent
susceptibility.
This condition together with the definition of the effective coupling
coupling constant \eqref{mcoupling} leads to
\beq
  \ln\left(\frac{\pi}{e^{\gamma_E}}\frac{T}{2^{1/3}\Delta_0}\right)%
  =\int_{0}^{\infty}dl_{\parallel}%
 \frac{1}{l_{\parallel}}\tanh\left(\frac{l_{\parallel}}{2T}%
 \right)\frac{3(1-\Phi)f_F(1-f_F)}{1-3(1-\Phi)f_F(1-f_F)}%
 \equiv{\mathcal F}(1-\Phi).
\eeq 
Note that the quantity ${\mathcal F}$ is dimensionless and does 
not depend on $T$. 
In the case of the deconfinement phase with $\Phi=1$, ${\mathcal F}$
vanishes so that it simply reproduces the standard expression for the
critical temperature 
$T_c^0=\frac{e^{\gamma_E}}{\pi}2^{1/3}\Delta_0=0.714\Delta_0$. 
It is now easy to imagine that the deviation of $\Phi$ from unity leads to 
the positive ${\mathcal F}$, and thus increases the critical 
temperature $T_c=0.714\Delta_0 e^{{\mathcal F}(1-\Phi)}$. 
To the first order in $(1-\Phi)$, ${\mathcal F}$ can be calculated as
\beq
 {\mathcal F}(1-\Phi)\sim \frac{21\zeta(3)}{4\pi^2}(1-\Phi)=0.64(1-\Phi).
\eeq
Thus to this order, the critical temperature is approximated by
\beq
  T_c=0.714\Delta_0\times e^{0.64(1-\Phi)},
\eeq 
near $\Phi\sim 1$. 
When $\Phi=0.4$ is substituted into the above formula, $T_c$ gets the factor of 
enhancement $e^{0.64(1-\Phi)}\sim1.5$. 
Although it is within the linear level, this value fairly
agrees with the numerically obtained factor, $1.8$. 
If we use the numerical value of ${\mathcal F}(1-0.4)=0.584$, the 
factor of enhancement becomes $1.79$; the agreement is perfect. 
Although unrealistic, at $\Phi=0$ (the confinement), 
the analytical evaluation is also possible. 
In this case we have ${\mathcal F}(1)=\ln3^{3/2}$ 
so that we have the factor $e^{\mathcal F}=3\Sqrt{3}=5.2$. 
This is the theoretical maximum of the critical temperature 
in the PNJL model at weak coupling.

\subsection{Effect of charge neutrality; the two-step hierarchical
    unlocking transition}
Let us now discuss charge neutrality at $m_s=0$. 
In Fig.~\ref{gapvsT}(b), we show the gaps and chemical potentials as a
function of $T$ calculated respecting the charge neutrality constraints. 
For comparison, we have shown $\Delta_3$ without neutrality (in
Fig.~\ref{gapvsT}(a)) by a dashed line. 
At a first glance, we notice that, even quantitatively, the charge
neutrality plays only a minor role at $m_s=0$.

However, several interesting remarks deserve a discussion here. 
(i) First, the charge neutrality conditions lift the degeneracy
$\Delta_1=\Delta_2$ 
away and open a small window for the dSC $(\Delta_2=0)$ phase between
the CFL and 2SC phases. 
(ii) Second, $\mu_3$ vanishes when the 2SC phase sets in. 
(iii) Lastly, $\mu_8$ does not vanish even when all the pairing melt and
the system goes into the unpaired quark matter as already discussed in
Sec.~\ref{resoneutrality} in the case of no pairing at all. 
Without finite $\mu_8$, the unpaired system inevitably has a 
finite $T_8$-color charge. 
The reason for (ii) is simple. 
The 2SC pairing preserves the $SU(2)_c$ symmetry intact or, in other
words, the 2SC gap has $SU(2)_c$ singlet structure and is transparent to
the $SU(2)_c$ charge. 
Therefore the system with no $T_3$-color charge should have $\mu_3=0$. 
One may think that a finite value of $\phi_3$ induces a finite $T_3$
charge in the system with $\mu_3=0$, but this is not correct. 
In fact, as we saw in the discussion in Sec.~\ref{resoneutrality},
restricting ourselves to the real part of $\Omega$, $n_r=n_g\ne n_b$ is
realized in the unpaired system inducing a $T_8$ charge but no $T_3$
charge. 
The point (i) can be understood as follows. 
In the presence of $\phi_3$, the gaps split as $\Delta_1=\Delta_2<\Delta_3$ 
as we saw above. 
This induces an imbalance in the thermal population of nine quasiquarks
in the CFL phase. 
Accordingly, the charge neutrality is lost unless $\mu_3$, $\mu_8$ and $\mu_e$ 
are tuned to their appropriate values. 
But of course the finite values of $\mu_e$ and $\mu_3$ explicitly break
the remaining $SU(2)_{c+V}$ symmetry.  
As a result of this secondary effect, the $SU(2)_{c+V}$ degeneracy
should be lifted away, as $\Delta_1\ne\Delta_2$. 
The appearance of the dSC phase at $m_s=0$ is in contrast either to the
Ginzburg-Landau approach \cite{Iida:2003cc} or to the pure 
NJL calculations \cite{Fukushima:2004zq}; this is definitely due to the
nontrivial interplay between the neutrality constraints and the Polyakov
loop dynamics.

\subsection{Effect of the stress due to nonzero strange quark mass}
\begin{figure}[tp]
  \includegraphics[width=0.457\textwidth,clip]{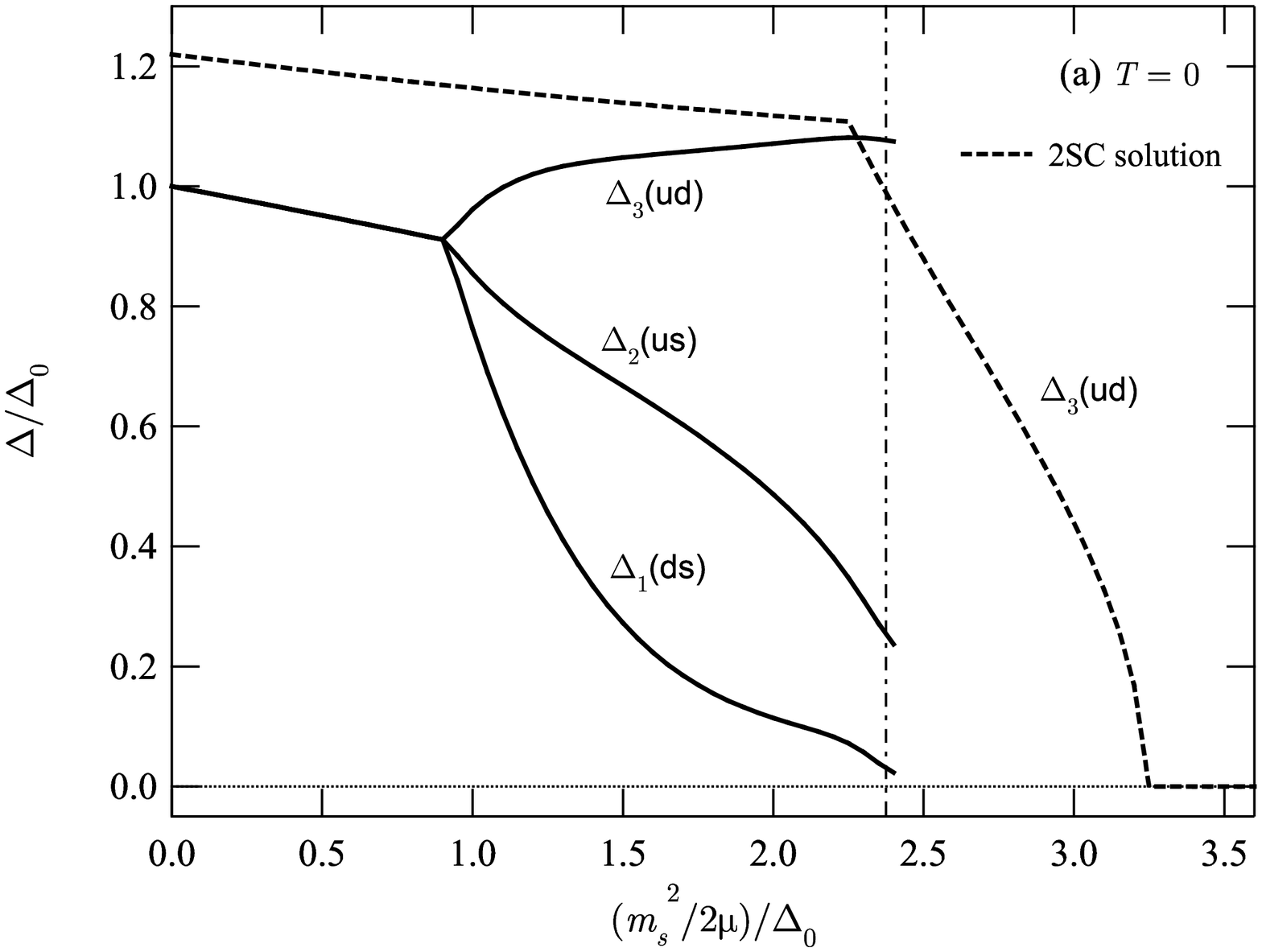}
  \includegraphics[width=0.457\textwidth,clip]{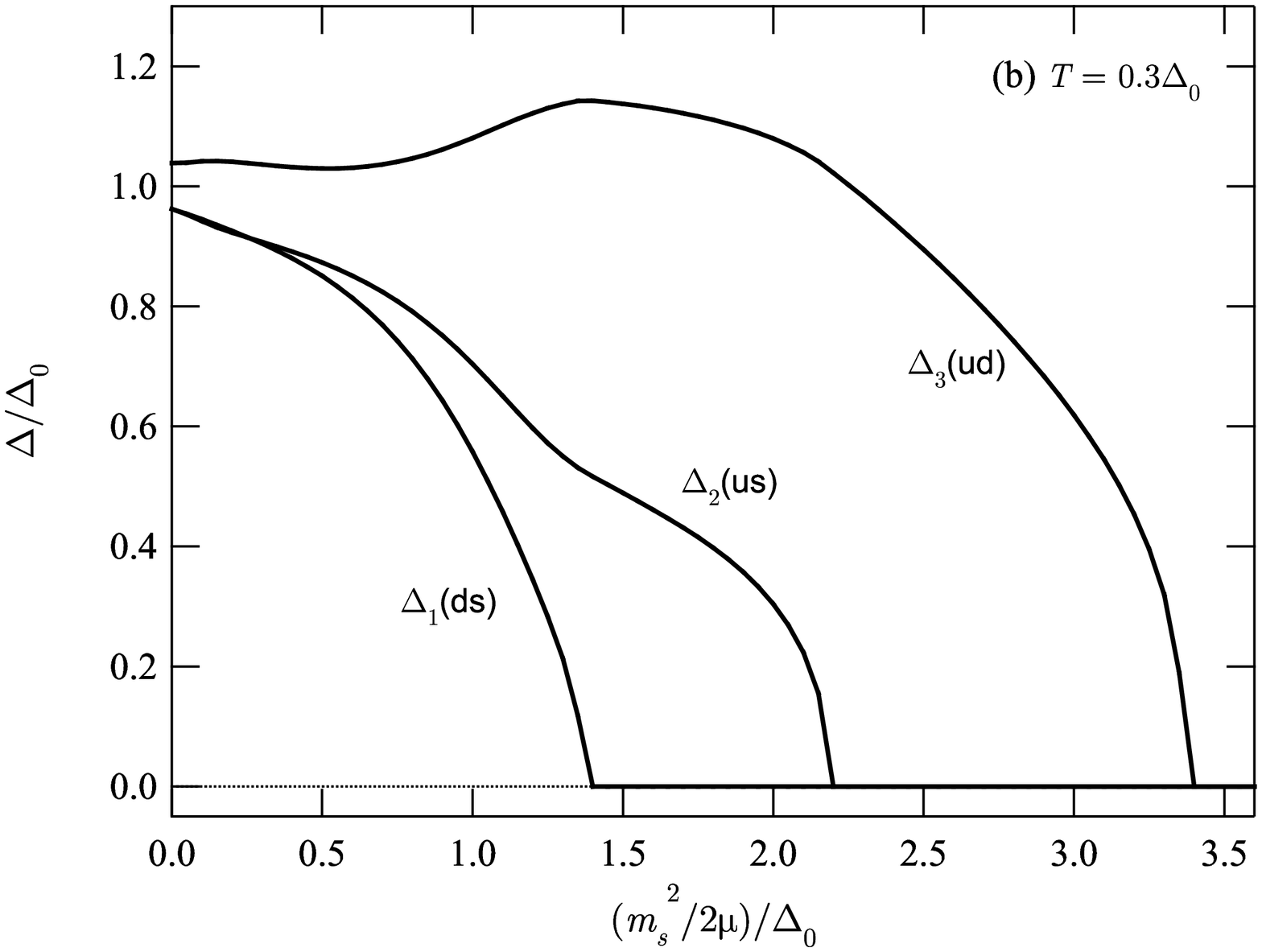}
 \caption[]{(a):~The gaps $\Delta_\eta$ as a function of $m_s^2/2\mu$
 at $T=0$. 
 The dashed line is $\Delta_3$ for the 2SC solution.
 The vertical dotted line corresponds to the 1st order gCFL-to-g2SC
 transition; on the left of it the CFL is realized, while on the right
 of it the 2SC is realized as the ground state.
 (b):~The same as (a) at finite temperature, $T=0.3\Delta_0$.
 All the transitions are of second order.
 }
 \label{gapvsms}
\end{figure}

Let us now discuss the effect of a nonzero strange quark mass on the
structure of gaps. 
In Fig.~\ref{gapvsms}(a) we show the zero temperature gaps $\Delta_\eta$
as functions of $\frac{m_s^2}{2\mu}$, both for the CFL (solid lines)
and 2SC (dashed line) solutions. 
The free energy comparison shows that there is a first order phase
transition from the gCFL phase to the g2SC phase at the point
$\frac{m_s^2}{2\mu}\cong 2.4\Delta_0$ 
indicated by the vertical dash-dotted line in the figure. 
Since at $T=0$ the Polyakov loop dynamics decouples from the pairing
(NJL) sector, the Polyakov loop plays no role in the gap structure. 
Consequently, the phase structure and the behavior of the gaps is similar 
to the result of \cite{Fukushima:2004zq} although the HDET approximation
was not adopted there.

At low $m_s$, the CFL phase is realized, and it continuously goes into
the gCFL phase \cite{Alford:2003fq} at a point slightly lower than
$\frac{m_s^2}{2\mu}=\Delta_0$. 
Then eventually the gCFL phase is taken over by the g2SC phase at
$\frac{m_s^2}{2\mu}\cong2.4\Delta_0$ 
by a first order transition \cite{Shovkovy:2003uu,Fukushima:2004zq}.
In Fig.~\ref{gapvsms}(b), we have shown the gaps
$\Delta_\eta(\frac{m_s^2}{2\mu})$ at a finite temperature, $T=0.3\Delta_0$. 
The first order transition is completely washed away, and there are two
successive second order unlocking transition until the system gets
unpaired, first from the CFL phase to the uSC phase, and subsequently
from the uSC phase to the 2SC phase. 
This feature is also qualitatively the same as in the NJL calculations
\cite{Fukushima:2004zq}.

\subsection{Interplay of the Polyakov loop dynamics
and enforced neutrality at finite strange quark mass}
\begin{figure}[tp]
  \includegraphics[width=0.457\textwidth,clip]{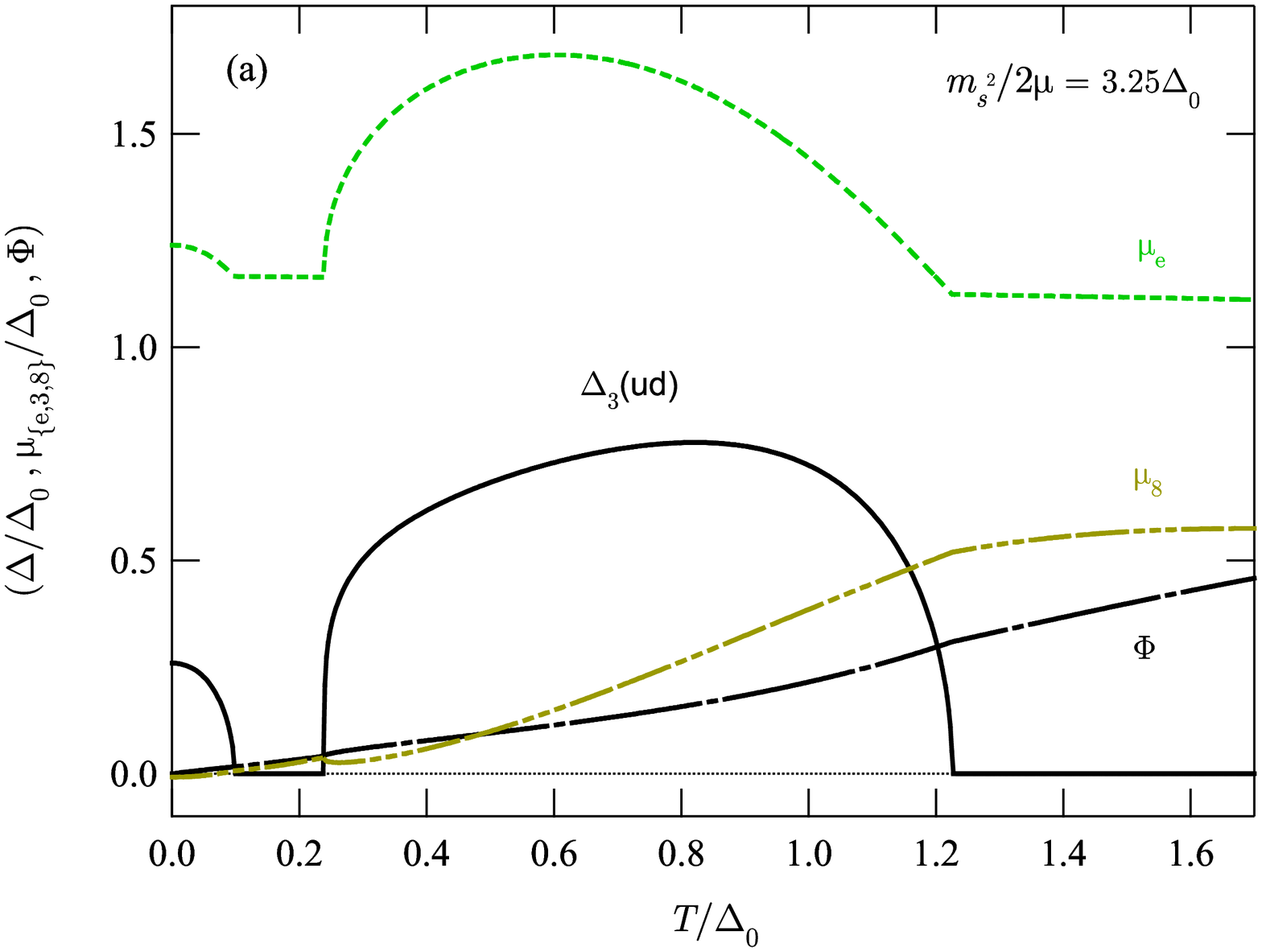}
  \includegraphics[width=0.457\textwidth,clip]{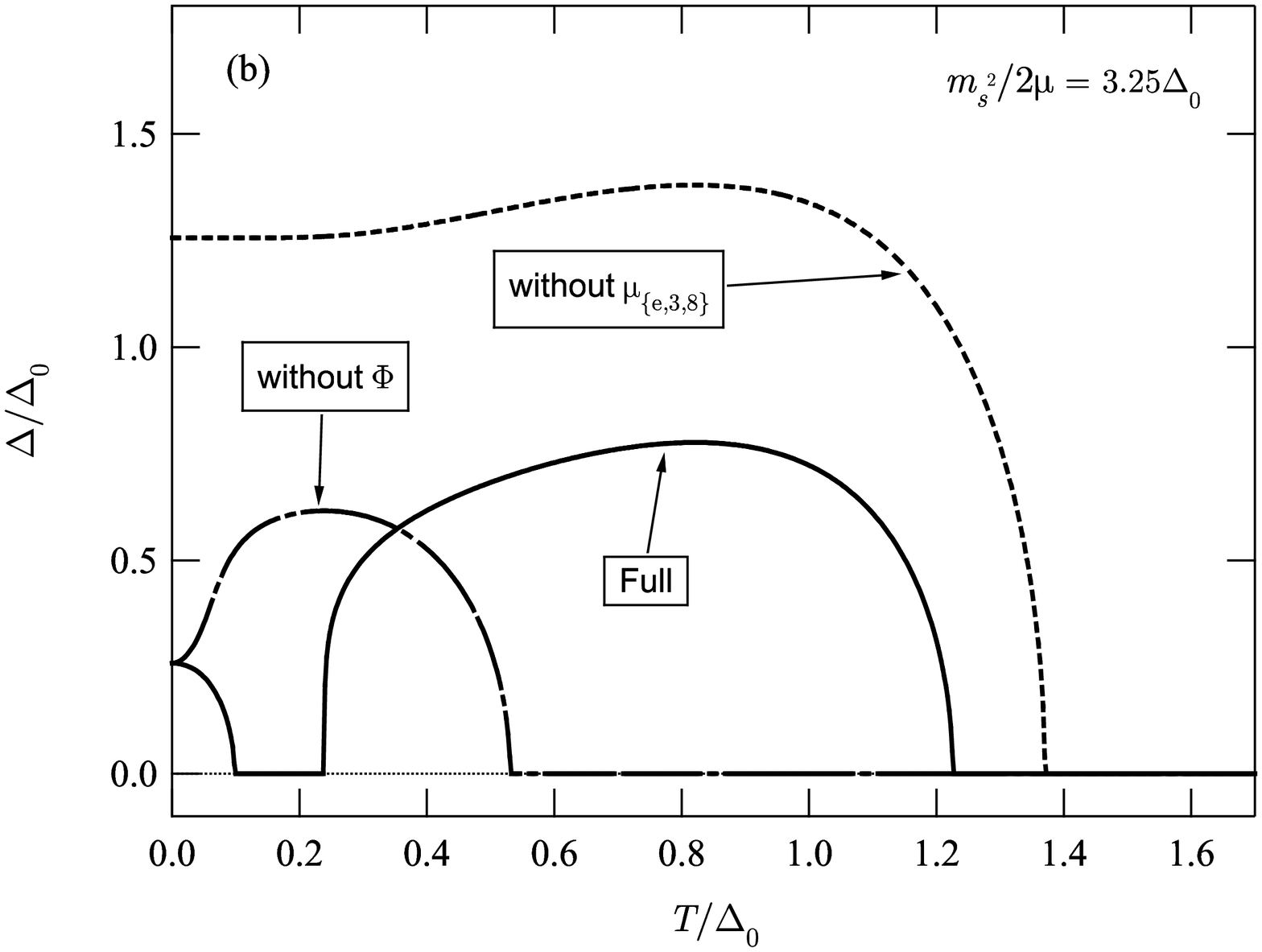}
 \caption[]{(a):~The gaps $\Delta_\eta$, the chemical potentials $\mu_{e,8}$,
 and the traced Polyakov loop $\Phi$ as a function of $T$
 at $m_s^2/2\mu=3.25\Delta_0$.
 (b):~Comparison of (a) with simplified versions of the model. 
 The bold full line is the same as $\Delta_3(T)$ in (a). 
 The long-dashed line is the 2SC gap $\Delta_3(T)$ calculated in the pure 
 NJL model without the Polyakov loop dynamics. 
 The dashed line is $\Delta_3(T)$ calculated with the Polyakov loop but
 without respecting charge neutralities, \ie, ${\mu_{e,3,8}}=0$.
 }
 \label{gapvsTms}
\end{figure}

Let us finally examine the impact of enforcing charge neutralities
and including the Polyakov loop dynamics into the pairing phases at an
intermediate density represented by the finite value of
$\frac{m_s^2}{2\mu}$. 
In Fig.~\ref{gapvsTms}(a) the gap $\Delta_3$ and the charge chemical
potentials $\{\mu_e,\mu_8\}$ are depicted as functions of $T$ at
$\frac{m_s^2}{2\mu}=3.25\Delta_0$. 
At this value of the stress, the CFL pairing is not possible, so only the 
(g)2SC phase can show up as a pairing pattern. 
What is surprising and also intriguing is that $\Delta_3$ once melts at
$T\cong0.1\Delta_0$ 
but appears again at higher temperature about $0.24\Delta_0$, 
and then finally vanishes completely when $T$ exceeds $1.23\Delta_0$. 
The 2SC phase exists in two different region in temperature. 
This feature is definitely due to the inclusion of both the neutrality
and the Polyakov loop into the problem. 
To see this, we show in Fig.~\ref{gapvsTms}(b), 
the gaps calculated with simplified versions of our PNJL model. 
The solid line indicated by ``Full'' is the same as $\Delta_3$
in Fig.~\ref{gapvsTms}(a). 
The long-dashed line represents the result calculated using the 
pure NJL model without the Polyakov loop dynamics, while the dashed line
is that calculated with PNJL model but without imposing the charge
neutrality constraints, \ie, putting $\mu_{e,3,8}=0$ from the very
beginning.
From these comparisons, it is clear that the appearance of the
intriguing possibility of the existence of two islands of 2SC in
temperature is due to the combinatory, cooperative effect between 
the Polyakov loop dynamics and the neutrality constraints. 
In contrast to the case with $m_s=0$, imposing neutrality has a sizable
effect on the gap. 
It significantly reduces the magnitude of the gap. 
It is so because in the case of a finite value of $m_s$, not only $\mu_8$ 
but also $\mu_e$ should be finite even in the unpaired phase in order to
guarantee electrical neutrality. 
Moreover the effect of the Polyakov loop is not only to stabilize the
2SC phase against the increase of temperature as in $m_s=0$ case, but
also to suppress the pairing at low temperature making two separated
islands of 2SC in temperature.

\section{Summary}\label{sum}
In this paper, we have studied the quark matter phase structure in
$\left(T,\frac{m_s^2}{2\mu}\right)$ 
plane starting from the PNJL model in which a temporal static gluon
field couples with quarks. 
This work is a natural extension of previous studies
\cite{Roessner:2006xn,Ciminale:2007ei,Ciminale:2007sr}. 
The particular focus was put on the effects produced by the inclusion of
the Polyakov loop dynamics on the pairing phases and by the enforcement
of color and electrical neutrality.

In the conventional PNJL model, there is a mismatch in each color
density so that the model lacks the color neutrality even in the
unpaired phase. 
This unphysical feature is significant in the proximity of the deconfinement 
transition. 
We have pointed out that this behavior may be due to the original
assumption hidden in the PNJL model \ie  that the traced Polyakov loop
dynamics can be represented by the inclusion of the static temporal
``colored'' gauge field which couples to the fundamental color charge of
dynamical quarks. 
By this assumption one misses gauge invariance. 
Once this fact is admitted, in order to avoid the unphysical appearance of 
color densities within this model, one has to include the charge
chemical potentials into the problem from the beginning. 
In fact, we have shown that $\mu_8$ should be finite to maintain color
neutrality in the unpaired phase.

In the detailed numerical analysis, we have depicted the phase 
diagram in $(T,m_s^2/2\mu)$-plane, and clarified how the phase 
diagram is affected by the inclusion of the Polyakov loop and the 
enforcement of charge neutrality. 
Even at $m_s=0$, the effect of the Polyakov loop is remarkable; it
breaks the $SU(3)_{c+V}$ down to $SU(2)_{c+V}$ and causes a continuous
color-flavor unlocking at finite temperature in a novel mechanism. 
In addition, it makes the 2SC phase much more robust against the
increase of temperature. 
The critical temperature is about twice as large as the weak coupling
prediction, which is consistent with previous calculations
\cite{Roessner:2006xn,Ciminale:2007ei,Ciminale:2007sr}. 
We have also examined a formal explanation about these facts and derived an 
analytical expression of this enhancement factor; it turned out that the
temporal gauge field reduces a blocking integral.

The effect of imposing neutralities gives only a tiny effect at $m_s=0$
although it opens a small window for the dSC realization between the 2SC
and CFL phases by lifting away the $SU(2)_{c+V}$ degeneracy.
In this case, we have a hierarchical unlocking, CFL~$\to$~dSC$\to$~2SC,
until it eventually goes into the unpaired phase.
This is contrast to the pure NJL calculation without the Polyakov loop
where the dSC never shows up at $m_s=0$ \cite{Fukushima:2004zq}.

The sizable effect of imposing charge neutrality on the pairing 
phases manifests itself at finite $m_s$. 
We have shown that the nontrivial, complicated interplay between the charge 
neutrality constraints and the Polyakov loop dynamics at $m_s\ne0$
produces a thermal reentrance phenomenon, as two isolated windows for the 
2SC pairing can show up on the temperature axis.

There are several ways to extend our current study. 
One is to take the chiral condensation into account by including 
the chiral condensate and removing the high density approximation
\cite{Abuki:2004zk}. 
By this improvement, one can study the interplay between the chiral
condensate, the Polyakov loop, and color superconductivity at the 
same time. 
The other possibility is to study mesonic modes 
\cite{Hansen:2006ee,Abuki:2008tx} 
as well as the Meissner masses in the superconducting phases. 
This might have an impact either on a possible meson condensation 
in superconducting phases or on the so called chromomagnetic instability
problem in gapless phases \cite{Casalbuoni:2004tb,Huang:2004bg}. 
These studies may be presented elsewhere in future.

To conclude let us stress that two alternatives have presented to us:

 a) application of the PNJL model to finite density is pathological and
 should be avoided, in relation to the fact that color neutrality is not
 visibly satisfied;

 b) the model can be used also at finite density provided neutrality is
 enforced: 
 for such a case we have derived the detailed consequences
 obtaining surprising results but without apparent physical
 inconsistencies.

 We hope that further work will illuminate on the choice between a) and b).

\vspace*{1ex} 
\noindent
We thank Kenji Fukushima, Mei Huang, Andreas Schmitt and in particular,
Igor Shovkovy for enlightening discussion.
One of us (H. A.) thanks I.N.F.N. for financial support. 
Numerical calculations were carried out on Altix3700 at YITP in Kyoto
University, and on the workstation {\small\sc NETCLUS} at University of Bari
(I.N.F.N., Sezione di Bari).

\end{document}